\documentclass[journal,10pt]{IEEEtran}
\usepackage{etoolbox}
\usepackage{dtsc-creafig}
\usepackage{notation}

\usepackage{epsfig}
\usepackage{color}
\usepackage{amsmath}
\usepackage{amssymb}
\usepackage{mathabx}
\usepackage{enumerate}
\usepackage{multirow}
\usepackage{bbm}
\usepackage{algorithm}
\usepackage{algorithmic}
\usepackage{lipsum,amsmath,multicol}
\usepackage{stfloats}
\usepackage{arydshln} 
\usepackage{amsfonts}
\usepackage{dsfont}
\usepackage{upgreek}
\usepackage{pgfplots} 
\usepackage{caption}
\usepackage{subfigure} 
\usepackage{booktabs} 

\usepackage{cite}

\definecolor{mycolor}{cmyk}{1,0,1,0}
\definecolor{mycolor1}{rgb}{1.00000,0.50000,0.00000}%
\definecolor{mycolor2}{rgb}{0.00000,0.80000,1.00000}%
\definecolor{mycolor3}{rgb}{1.00000,0.00000,1.00000}%
\definecolor{mycolor4}{rgb}{0.45, 0.31, 0.59}%
\definecolor{mycolor5}{rgb}{0.6, 0.4, 0.8}
\definecolor{carnationpink}{rgb}{1.0, 0.65, 0.79}
\definecolor{auburn}{rgb}{0.43, 0.21, 0.1}

\usepackage{acronym}
\acrodef{LMMSE}{linear minimum mean square error}
\acrodef{FIR}{finite impulse response}
\acrodef{MIMO}{multiple-input multiple-output}
\acrodef{i.i.d.}{independent and identically distributed} 
\acrodef{S/P}{serial to paralell} 
\acrodef{BP}{belief propagation}
\acrodef{SPA}{sum-product algorithm}     
\acrodef{GMP}{Gaussian message passing} 
\acrodef{KL}{Kullback-Leibler} 
\acrodef{ML}{maximum likelihood}
\acrodef{pdf}{probability density function}
\acrodef{SISO}{single-input single-output}
\acrodef{DFE}{decision feedback equalization}
\acrodef{EXIT}{extrinsic information transfer}
\acrodef{BER}{bit error rate}
\acrodef{SER}{symbol error rate}
\acrodef{ECC}{error correction code}
\acrodef{FEC}{forward error correction}
\acrodef{MSE}{mean-square-error}
\acrodef{MMSE}{minimum mean square error}
\acrodef{TLMMSE}{turbo linear minimum mean square error}
\acrodef{APP}{a posteriori probabilities}
\acrodef{LLRs}{log-likelihood ratios}
\acrodef{LLR}{log-likelihood ratio}
\acrodef{BEP}{block expectation propagation}
\acrodef{nuBEP}{non-uniform block expectation propagation}
\acrodef{TBEP}{turbo block expectation propagation}
\acrodef{ISI}{intersymbol interference}
\acrodef{ICI}{intercarrier interference}
\acrodef{AWGN}{additive white Gaussian noise}
\acrodef{MAP}{maximum a posteriori}
\acrodef{LDPC}{low-density parity-check}
\acrodef{EP}{expectation propagation}
\acrodef{SD}{sphere decoding}
\acrodef{MCMC}{Markov chain Monte Carlo}
\acrodef{GTA}{Gaussian tree approximation}
\acrodef{CHEMP}{channel hardening-exploiting message passing}
\acrodef{BP}{belief propagation}
\acrodef{LTI}{linear time invariant}
\acrodef{pmf}{probability mass functions}
\acrodef{KL}{Kullback-Leibler}
\acrodef{KSEP}{Kalman smoothing expectation propagation}
\acrodef{SEP}{smoothing expectation propagation}
\acrodef{BP-EP}{belief propagation expectation propagation}

\ifCLASSINFOpdf
\else
\fi
\begin{document}
%
\title{Equalization with Expectation Propagation at Smoothing Level}
%
%
%

\author{Irene~Santos, 
        Juan~Jos\'e~Murillo-Fuentes and Eva~Arias-de-Reyna
\thanks{I. Santos, J.J. Murillo-Fuentes and E. Arias-de-Reyna are with the Dept. Teor\'ia de la Se\~nal y Comunicaciones, Universidad de Sevilla, Camino de los Descubrimientos s/n, 41092 Sevilla, Spain. E-mail: {\tt \{irenesantos,murillo,earias\}@us.es}}
\thanks{This work was partially funded by Spanish government (Ministerio de Econom\'ia y Competitividad TEC2016-78434-C3-R) and the European Union (FEDER).}}

\maketitle

\begin{abstract}

In this paper we propose a smoothing turbo equalizer based on the expectation propagation (EP) algorithm with quite improved performance compared to the  Kalman smoother, at similar complexity. 
%
In scenarios where high-order modulations or/and large memory channels are employed, the optimal BCJR algorithm is computationally unfeasible. In this situation, low-cost but suboptimal solutions, such as the \ac{LMMSE}, are commonly used. Recently, EP has been proposed as a tool to improve the Kalman smoothing performance.  
 In this paper we review these solutions to apply the EP at the smoothing level, rather than at the forward and backwards stages. Also, we better exploit the information coming from the channel decoder in the turbo equalization schemes. With these improvements we reduce the computational complexity, speed up convergence and outperform previous approaches. We included some simulation results to show the robust behavior of the proposed method regardless of the scenario, and its improvement in terms of performance in comparison with other EP-based solutions in the literature.

%
%


\end{abstract}


%
\IEEEpeerreviewmaketitle

\section{Introduction}


The task of a soft equalizer is to mitigate the \ac{ISI} introduced by the channel, providing a probabilistic estimation of the transmitted symbols given a set of observations. After the equalizer, a channel decoder can help to detect and correct errors if the transmitted message has been protected with some redundancy \cite{Proakis08}. These channel decoders highly benefit from soft estimations provided by the equalizer rather than hard decisions. In addition, the equalization can be refined with the help of the information at the output of the channel decoder with turbo equalization \cite{Douillard95,Berrou10,Tuchler11}. 

The BCJR provides a \ac{MAP} estimation \cite{Bahl74}. However, the BCJR solution becomes intractable in terms of complexity as the memory and/or the constellation size grow. In this scenario, some approximated BCJR solutions, such as \cite{franz98,Colavolpe01,Sikora05,Fertonani07}, can be employed. These solutions reduce the complexity by just exploring a reduced number of states. However, their performance is quite dependent on the channel realization and they degrade if the number of surviving states does not grow accordingly with the complexity of the scenario. 
A comparison between them can be found in \cite{Santos16,Santos17}. 

A different and extended approximate solution is the linear minimum mean square error (\ac{LMMSE}). An equalizer based on the LMMSE algorithm can be implemented, among others, as a block \cite{Muranov10}, a Wiener-filter type \cite{Tuchler11,Koetter02,Singer01,Koetter04} or a Kalman smoothing \cite{Park15} approach. 
%
%
The Kalman smoothing solution exhibits the performance of the block \ac{LMMSE} with linear complexity in the length of the transmitted word and quadratic in the memory of the channel. It proceeds \emph{forwards} by computing the transition probabilities from consecutive states through a trellis representation. Then, the same procedure is run \emph{backwards}. Finally, it merges both procedures into a \emph{smoothing} approach, providing an approximation to the posterior of each transmitted symbol.

However, the LMMSE estimation is far from optimal. It is a linear solution in the observation, where the discrete priors
are approximated by Gaussian distributions whose mean and variance are set to the ones of the discrete distribution at the output of the channel decoder. When this information is not available, they are set to zero mean Gaussian distributions of variance equal to the energy of the constellation. 
%
The \ac{EP} algorithm exhibits a structure similar to that of the LMMSE, but the \notaI{priors} depend on the observation, hence being non-linear. \notaIb{The set of Gaussian distributions used to approximate the priors is estimated iteratively, with the aim of better approximating the posterior distribution of the transmitted symbols.}
%
%
The \ac{EP} computes the mean and variance of these Gaussians by matching the moments of the approximated posterior with the ones of the true posterior, including the discrete \ac{pmf} of the priors. 
%
 The \ac{EP} has already been applied to \ac{MIMO} detection \cite{Cespedes17,Cespedes14,Santos18b,Senst11}, \ac{LDPC} channel decoding \cite{Olmos13b,Salamanca13} and standalone/turbo equalization \cite{Santos17,Santos16,Santos18,Sun15,Hu06}, among others. 

In \cite{Santos16}, a block implementation of an EP-based equalizer is proposed, whose complexity is quadratic in the frame length. This implementation is revised in \cite{Santos18} to propose an optimized version for turbo equalization. Since the block implementation can be intractable for long frames, a filtered implementation based on the Wiener-filter behavior is also proposed in \cite{Santos18} for turbo equalization. It allows a reduction in complexity which is linear in the frame length and quadratic in the size of the window used. However, the Wiener-filter implementation only uses the observations within the window to obtain an estimation for the transmitted symbols, which can degrade the performance in comparison with its block counterpart. To avoid the inversion of large matrices that are needed in the block EP-based solution without degrading the performance as in the Wiener-filter proposal, a smoothing equalizer based on the EP algorithm can be employed. 

In this framework of smoothing equalizers, in \cite{Sun15,Hu06} the \ac{EP} was applied to better exploit the information fed back from the channel decoder in turbo equalization. However, this approach degrades when used for higher order modulations  since no suitable control of negative variances is used, a minimum allowed variance is not set and no damping procedure is included. \notaIb{Also, this equalizer is just designed for turbo equalization, boiling down to the LMMSE for standalone equalization. }
%
%
In \cite{Santos17} the authors proposed a self-iterated \ac{EP} equalizer, the \ac{SEP}, that improves the performance either as standalone or turbo equalization. 
The \ac{EP} was introduced to improve the estimation of the posteriors in the forward and backward approaches. Then, both forward and backward approximations were merged into a smoothing distribution. To get a performance similar to that of the block \ac{EP} equalizer \cite{Santos16}, a state involving a number of symbols larger\footnote{Approximately twice the channel length.} than the channel length was needed. 

\notaIb{In this paper we propose a new Kalman smoothing \ac{EP} equalizer where we have introduced the following improvements. First, the \ac{EP} is introduced at the smoothing step, improving the convergence in comparison to \cite{Santos17}. Second, the true prior used in the moment matching procedure of the \ac{EP} algorithm is set to a non-uniform distribution provided by the channel decoder \cite{Santos18}. This is a further difference with respect to the \ac{SEP} approach in \cite{Santos17}, where uniform priors were used even after the turbo procedure. \notaI{Third, }a better control of the minimum allowed variance at the moment matching is introduced. 
Finally, we review the forgetting factor used at the damping procedure. As a result, better gains in terms of \ac{BER} are achieved, the convergence in turbo equalization is more robust compared to the one of the equalizers in \cite{Sun15,Santos17} and the computational complexity yields quadratic in the channel length, i.e., the one of the Kalman smoothing solution.}


The paper is organized as follows. In \SEC{sysmod} we describe the model of the communication system with feedback from the channel decoder. \SEC{smoothingLMMSE} is devoted to reviewing the LMMSE from a smoothing point of view, since it will be the starting point for our proposal. Then, in \SEC{EP}, we describe the novel smoothing EP, which is applied at the smoothing stage. In \SEC{results}, we include some simulation results to show the robustness of the proposal and compare it to the LMMSE and other EP-based solutions. We end with conclusions in \SEC{conc}. 

We use the following specific notation throughout the paper. We denote the $i$-th entry of a vector $\vect{\beforechannel}$ as 
$\beforechannel_i$, its entries in the range $[i, j]$ as $\vect{\beforechannel}_{i:j}$  and its Hermitian transpose as $\matr{u}\her$. We use $\cgauss{\vect{\beforechannel}}{\boldsymbol{\upmu}}{\boldsymbol{\Sigma}}$ to denote a normal distribution of a random proper complex vector 
$\vect{\beforechannel}$ with mean vector $\boldsymbol{\upmu}$ and covariance  matrix $\boldsymbol{\Sigma}$.  The expression $\mbox{Proj}_G[q(\cdot)]=\gauss{\cdot}{m}{\sigma^2}$ is the projection of the distribution given as an argument, $q(\cdot)$, with moments $m$ and $\sigma^2$, respectively, into the family of Gaussians \cite{Sun15}.

\section{System Model}\LABSEC{sysmod}

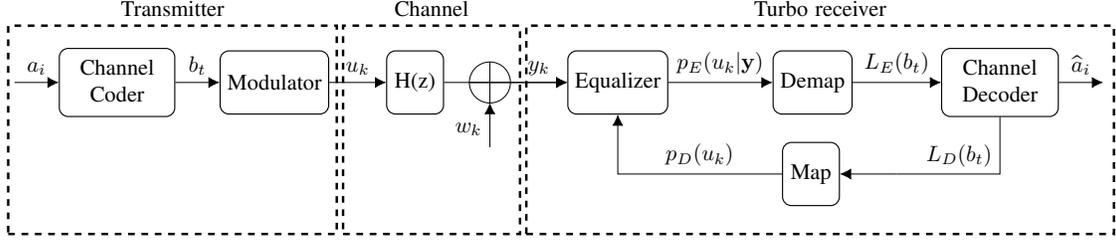
\begin{figure*}[htb]
\centering
\scalebox{0.84}{\tikzset{block/.style = {draw, rectangle, line width=0.5pt,
minimum height=1cm, minimum width=0.9cm,rounded corners},
input/.style = {coordinate},
output/.style = {coordinate},
window/.style={rectangle,draw,dashed,minimum width=6.8cm,minimum height=3.3cm,line width=1.1pt},
virtual/.style={coordinate},
}

\begin{tikzpicture}[auto, >={Latex[width=5pt,length=5pt]}]
\node [input, name=input] {};
\node [block, right = 0.7 cm of input, name=cod] {\begin{tabular}{c}Channel\\Coder\end{tabular}};
\node [block, right = 0.7 cm of cod, name=mod] {Modulator};
\node [block, right = 0.9 cm of mod, name=channel] {H(z)};
\node [misuma, right=0.4 cm of channel, minimum size=0.65cm ] (sumador) {};     
\node [block, right =0.9 cm of sumador, name=eq] {Equalizer};
\node [block, right =1.65 cm of eq, name=demap] {Demap};
\node [virtual, right =0.7 cm of demap, name=v1] {};
\node [block, right = 0.7 cm of v1, name=decoder] {\begin{tabular}{c}Channel\\Decoder\end{tabular}};
\node [input, below =0.7 cm of sumador] (noise) {};
\node [output, right = 0.7 cm of decoder] (output) {};
\node[virtual, below=1.45 cm of v1, name=sum2] {};
\node [block, left = 0.9 cm of sum2, name=map] {Map};
\draw [->] (input) -- node {$\displaystyle a_i$} (cod);
\draw [->] (cod) -- node {$\displaystyle b_t$} (mod);
\draw [->] (mod) -- node {$\displaystyle \beforechannel_\iter$} (channel);
\draw [->] (channel) -- node {} (eq);
\draw [->] (noise) -- node[pos=0.4,left] {$\displaystyle \noise_\iter$} (sumador);
\draw [->] (sumador) -- node {$\displaystyle \afterchannel_\iter$} (eq);
\draw [->] (eq) -- node {$\displaystyle p_E(\beforechannel_\iter|\vect{\afterchannel})$} (demap);
\draw [->] (demap) -- node {$\displaystyle L_E(b_t)$} (decoder);
\draw [->] (decoder) -- node {$\displaystyle \widehat{a}_i$}(output);
\draw [-] (decoder.south) |- node[left,pos=0.34] {$\displaystyle L_D(b_t)$}(sum2);
\draw [->] (sum2) -- node[above] {} (map);
\draw [->] (map) -| node[above,pos=0.25] {$\displaystyle p_{D}(\beforechannel_\iter)$}(eq); 

\node[window, name=win1,minimum width=5.2cm,label=above:Transmitter] at (2.5,-0.75) {};
\node[window, name=win1,minimum width=2.8cm,label=above:Channel] at (6.6,-0.75) {};
\node[window, name=win1,minimum width=9.3cm,label=above:Turbo receiver] at (12.75,-0.75) {};

\end{tikzpicture}}
\caption{\small System model.} \LABFIG{sysmod}
\end{figure*} 

The model of a turbo communication system is depicted in \FIG{sysmod}. The information bit vector, $\vect{a}=[\allvect{a}{1}{\k}]\trs$ with $a_i\in\{0,1\}$, is encoded into the coded vector $\vect{b}=[\allvect{b}{1}{\t}]\trs$. This codeword is modulated into the vector of symbols $\vect{\beforechannel}=[\allvect{\beforechannel}{1}{{\tamframe}}]\trs$, where each symbol belongs to a complex $\modsize$-ary constellation with alphabet $\mathcal{A}$ and mean transmitted symbol energy $\energy$. This vector of symbols is transmitted over a channel with weights $\vect{h}=[\allvect{h}{\ntaps}{1}]$ and memory $\memch=\ntaps-1$ and it is corrupted with \ac{AWGN} with known noise variance, $\sigma_\noise^2$. The received signal is denoted as $\vect{\afterchannel}=[\allvect{\afterchannel}{1}{{\tamframe+\ntaps-1}}]\trs$ with entries given by
\begin{equation}\LABEQ{yk}
\afterchannel_\iter=\sum_{j=1}^{\ntaps} h_{j}\beforechannel_{\iter-j+1} + \noise_\iter= \vect{h}\trs\vect{\state}_k+\noise_\iter,
\end{equation}
where ${\noise_\iter}\sim\cgauss{{\noise_\iter}}{{0}}{\sigma_\noise^2}$
, $\vect{s}_k=[\beforechannel_{\iter-\memch},...,\beforechannel_\iter]\trs$ which will be hereafter referred to as the \emph{state} and $\beforechannel_\iter=0$ for $\iter<1$ and $\iter>\tamframe$. 
The received signal is processed by the equalizer. It estimates the posterior probability of the transmitted symbol vector $\vect{\beforechannel}$ given the whole vector of observations $\vect{\afterchannel}$ as
\begin{equation}\LABEQ{pugiveny}
p(\vect{\beforechannel}|\vect{\afterchannel})\; \propto \;{p(\vect{\afterchannel}|\vect{\beforechannel})\prod\limits_{\iter=1}^{\tamframe} p(\beforechannel_\iter)} 
\end{equation}
where $p(\beforechannel_\iter)$ is the available information on the priors. If the output of the channel decoder is available and fed back to the equalizer, then $p(\beforechannel_\iter)=p_D(\beforechannel_\iter)$. Otherwise, a uniform discrete prior is used, which is equivalent to assuming equiprobable symbols. 
%
 The equalizer and the channel decoder usually exchange extrinsic information. An extrinsic distribution, $p_E(\beforechannel_\iter|\vect{\afterchannel})$, is computed at the output of the equalizer. The extrinsic distributions are demapped and given to the channel decoder as extrinsic log-likelihood ratios (LLRs), $L_E(b_t)$. Finally, the decoder computes extrinsic LLRs, $L_D(b_t)$, which are mapped again onto the $\modsize$-ary modulation and fed back to the equalizer.

%

\section{From the BCJR to the LMMSE}\LABSEC{smoothingLMMSE}
\subsection{BCJR}
%
\notaIb{In this section, we review the formulation for exact inference in equalization from a Kalman smoothing point of view.}
\notaI{In the forward stage of the BCJR}, at step $\iter$ the posterior probability distribution of the current state, $\vect{\state}_\iter$, is computed 
%
 given the observations up to time $\iter$, $\vect{\afterchannel}_{1:\iter}$. This posterior, $p(\vect{\state}_\iter|\vect{\afterchannel}_{1:\iter})$, is proportional to the product of the Gaussian likelihood of the current observation, $p(\afterchannel_\iter|\vect{\state}_\iter)$, and the predicted state, $p(\vect{\state}_\iter|\vect{\afterchannel}_{1:\iter-1})$, i.e., 
\begin{equation}\LABEQ{posState}
p(\vect{\state}_\iter|\vect{\afterchannel}_{1:\iter})\;\propto\; p(\afterchannel_\iter|\vect{\state}_\iter) \underbrace{p_D(\beforechannel_\iter)p(\vect{\state}_{\iter-1}^{\backslash 1}|\vect{\afterchannel}_{1:\iter-1})}_{p(\vect{\state}_\iter|\vect{\afterchannel}_{1:\iter-1})}
\end{equation}
where $p(\afterchannel_\iter|\vect{\state}_\iter)$ is given by the channel model in \EQ{yk} and 
$p(\vect{\state}_{\iter-1}^{\backslash 1}|\vect{\afterchannel}_{1:\iter-1})$ denotes the marginal of $p(\vect{\state}_{\iter-1}|\vect{\afterchannel}_{1:\iter-1})$ over its first entry, i.e., over $\beforechannel_{\iter-\ntaps}$. 
%
Similarly, and in parallel, a backward procedure can be run following the same formulation explained above by just left-right flipping the channel, received and transmitted vectors \cite{Santos17}. As a result, the distribution $p(\vect{\state}_\iter|\vect{\afterchannel}_{\iter:\tamframe+\memch})$ is estimated. 
The BCJR approach computes the posterior using the results of these forward and backward steps, to later merge the information, see Appendix, as: 
\begin{align}\LABEQ{trueSmooth1}
p(\vect{\state}_k|\vect{\afterchannel}) \;\propto\; \frac{p(\vect{\state}_k|\vect{\afterchannel}_{1:\iter})p(\vect{\state}_k|\vect{\afterchannel}_{\iter:\tamframe+\memch})}{p(\afterchannel_\iter|\vect{\state}_\iter)\prod\limits_{\iterw=\iter-\memch}^{\iter}p(\beforechannel_\iterw)}.
\end{align}
The computation of these probabilities has complexity $\order(\modsize^{\ntaps})$. For large states, i.e., high $\ntaps$, and/or large constellations sizes, $\modsize$, it becomes unaffordable due to the large number of combinations to be checked to retain the maximum value. 

\subsection{Kalman Smoother}
Rather than computing the true distribution in \EQ{trueSmooth1}, in the smoothing Kalman filtering we approximate $p(\vect{\state}_k|\vect{\afterchannel}_{1:\iter})$, $p(\vect{\state}_k|\vect{\afterchannel}_{\iter:\tamframe+\memch})$ and $p(\beforechannel_\iterw)$ by Gaussian distributions, as follows. 

The LMMSE approximates the discrete true prior, $p(\beforechannel_\iter)$, with a Gaussian distribution, ${t_\iter}(\beforechannel_\iter)$, as
\begin{equation}\LABEQ{aproxpriorMMSE}
{t_\iter}(\beforechannel_\iter)= \mbox{Proj}_G[p(\beforechannel_\iter)]\sim \cgauss{\beforechannel_\iter}{\mu_{t_\iter}}{\sigma_{t_\iter}^2}.
\end{equation}
If no feedback is available from the channel decoder, these moments are initialized to $\mu_{t_\iter}=0$ and $\sigma_{t_\iter}^2=E_s$.

The approximation above yields 
the following Gaussian approximation of the posterior in \EQ{posState}
\begin{equation}\LABEQ{posStateaprox}
{q^F}(\vect{\state}_\iter)\;\propto\; p(\afterchannel_\iter|\vect{\state}_\iter){t_\iter}(\beforechannel_\iter) {q^F}(\vect{\state}_{\iter-1}^{\backslash 1})\sim\cgauss{\vect{\state}_\iter}{\boldsymbol{\upmu}_\iter^F}{\vect{\Sigma}_\iter^F}
\end{equation}
where 
\begin{align}
\boldsymbol{\upmu}_\iter^F&=\matr{\Sigma}_\iter^F\left(
\std{\noise}^{-2}\vect{h}\her\afterchannel_{\iter}
+ 
\begin{bmatrix}
\left(\matr{\Sigma}^{\backslash 1}_{\iter-1}\right)\inv\boldsymbol{\upmu}^{\backslash 1}_{\iter-1}
 \\ \mu_{t_\iter}/\sigma_{t_\iter}^2
\end{bmatrix} 
\right), \LABEQ{meanqKEP}\\
\matr{\Sigma}_\iter^F&=\left(\std{\noise}^{-2}\vect{h}\her\vect{h}
+
\begin{bmatrix}
\left(\matr{\Sigma}_{\iter-1}^{\backslash 1}\right)\inv
&
\matr{0}_{(\ntaps-1)\times 1}\\ 
\matr{0}_{1\times (\ntaps-1)} & 1/\sigma_{t_\iter}^2
\end{bmatrix} 
\right)\inv \LABEQ{covqKEP}
\end{align}
 and the superscript $^F$ denotes the forward procedure. Vector $\boldsymbol{\upmu}_{\iter-1}^{\backslash 1}$ is 
 $\boldsymbol{\upmu}_{\iter-1}^F$
 without its first entry and $\vect{\Sigma}_{\iter-1}^{\backslash 1}$ is the submatrix  defined as $\matr{\Sigma}_{\iter-1}^F$ with its first row and column removed. 
This process is repeated for $\iter=1,...,\tamframe+\memch$. 
%
Note that the forward recursion only uses the observations $\afterchannel_1, ..., \afterchannel_\iter$  when estimating the posterior distribution in \EQ{posStateaprox}, ignoring the following $\afterchannel_{\iter+1}, ..., \afterchannel_{\tamframe+\memch}$. The computational complexity of this procedure is of $\order(\tamframe\ntaps^2)$ (see \cite{Sun15}). 


%


In the backwards step, the distribution $p(\vect{\state}_\iter|\vect{\afterchannel}_{\iter:\tamframe+\memch})$ is approximated by the following posterior Gaussian distribution, 
%
%
\begin{equation}\LABEQ{posStateaproxB}
{q^B}(\vect{\state}_\iter)\sim\cgauss{\vect{\state}_\iter}{\boldsymbol{\upmu}_\iter^B}{\vect{\Sigma}_\iter^B},
\end{equation}
%
where note that $\afterchannel_1, ..., \afterchannel_{\iter-1}$ are not involved. 

Finally, in the smoothing step both approximations in \EQ{posStateaprox} and \EQ{posStateaproxB} are merged  into just one posterior approximation of \EQ{trueSmooth1} as
\begin{equation}\LABEQ{KEPpost}
{q}(\vect{\state}_{\iter})=\frac{q^{F}(\vect{\state}_{\iter})q^{B}(\vect{\state}_{\iter})}{p(\afterchannel_\iter|\vect{\state}_{\iter})\prod\limits_{i=\iter-\memch}^\iter {t_i}(\beforechannel_i)}\sim\cgauss{\vect{\state}_{\iter}}{\boldsymbol{\upmu}_\iter}{\matr{\Sigma}_\iter}
\end{equation}
where 
\begin{align}\LABEQ{meanFB}
\boldsymbol{\upmu}_\iter&=\matr{\Sigma}_\iter \Big(\left(\matr{\Sigma}_{\iter}^{F}\right)\inv\boldsymbol{\upmu}_{\iter}^{{F}}+\left(\matr{\Sigma}_{\iter}^{B}\right)\inv\boldsymbol{\upmu}_{\iter}^{{B}} - \nonumber\\
&\hspace{0.4cm}-\std{\noise}^{-2}\vect{h}\her\afterchannel_{\iter} -\matr{C}_{{t_\iter}}\inv\boldsymbol{\mu}_{t_\iter} \Big), \\
\matr{\Sigma}_\iter&=\left( \left(\matr{\Sigma}_{\iter}^{F}\right)\inv+ \left(\matr{\Sigma}_{\iter}^{B}\right)\inv-\std{\noise}^{-2}\vect{h}\her\vect{h}-\matr{C}_{{t_\iter}}\inv \right)\inv, \LABEQ{sigmaFB}
\end{align}
$\boldsymbol{\mu}_{t_\iter}=[\mu_{t_{\iter-\memch}},...,\mu_{t_\iter}]\trs$ and $\matr{C}_{{t_\iter}}=\mbox{diag}([\sigma_{t_{\iter-\memch}}^2,...,\sigma_{t_\iter}^2])$. \notaIb{The computational complexity of this smoothing step is given by $\order(\tamframe\ntaps^2)$ (see \cite{Sun15}). Hence, the final complexity of the algorithm is $\order(2\tamframe\ntaps^2)$, i.e., the computational complexity of the forward/backward steps (that can be run in parallel) and the one of the smoothing step. }
%


In turbo equalization, the probabilities at the output of the channel decoder, $p_D(\beforechannel_\iter)$, are used as priors, $p(\beforechannel_\iter)$. Also, we usually handle the extrinsic probabilities to the channel decoder, which can be obtained as 
\begin{equation}\LABEQ{extrinsic}
q_E(\beforechannel_\iter)=\frac{{q}(\beforechannel_\iter)}{{t_\iter}(\beforechannel_\iter)}\sim\cgauss{\beforechannel_\iter}{{\mu_{E_\iter}}}{{\sigma_{E_\iter}^2}}
\end{equation}
where $q(\beforechannel_\iter)\sim\cgauss{\beforechannel_\iter}{\mu_{\iter}}{\sigma_\iter^{2}}$ is the marginal of \EQ{KEPpost} and 
%
\begin{align}
{\mu_{E_\iter}}&=\frac{\mu_{\iter}{\sigma}_{t_\iter}^2-\mu_{t_\iter} \sigma_\iter^{2}}{{\sigma}_{t_\iter}^2-\sigma_\iter^{2}}, \LABEQ{meanEXTgeneral}\\
{\sigma_{E_\iter}^2}&=\frac{\sigma_\iter^{2}{\sigma}_{t_\iter}^2}{{\sigma}_{t_\iter}^2-\sigma_\iter^{2}}.  \LABEQ{varEXTgeneral}
\end{align}
The whole procedure is summarized in \ALG{SmoothLMMSE}. 

\begin{algorithm}[!tb]
\begin{algorithmic}
\STATE 
{\bf Inputs}: ${t_\iter}(\beforechannel_\iter)$ for $\iter=1,...,\tamframe$
and $\afterchannel_\iter$ for $\iter=1,...,\tamframe+\memch$ 
\FOR {$\iter=1,...,\tamframe+\memch$ } 
\STATE
1) Compute the forward distribution, ${q^F}(\vect{\state}_\iter)$, in \EQ{posStateaprox}. 
\STATE
2)  Compute the backward distribution, ${q^B}(\vect{\state}_{\iter})$, in \EQ{posStateaproxB}. 
\ENDFOR 
\FOR {$\iter=1,...,\tamframe$}
\STATE
3) Compute the smoothed $\iter$-th distribution, ${q}(\vect{\state}_{\iter})$ in \EQ{KEPpost}. \\
\ENDFOR \\
4) Compute the marginals ${q}(\beforechannel_\iter)$.\\
5) Compute the extrinsics $q_E(\beforechannel_\iter)$ as in \EQ{extrinsic}. 
\STATE
{\bf Output}: Deliver $q_{E}(\beforechannel_\iter)$ to the decoder for $\iter=1,...,\tamframe$
\end{algorithmic}
\caption{Kalman Smoother Equalizer}\LABALG{SmoothLMMSE}
\end{algorithm}

\section{EP at smoothing level}\LABSEC{EP}

The EP \cite{Minka01thesis,Minka01,Seeger05,Bishop06} is a Bayesian machine learning technique to approximate an intractable \ac{pdf}, such as \EQ{pugiveny}, with an exponential family. In this paper we use it to estimate the factors, ${t_\iter}(\beforechannel_\iter)$, that minimize the \ac{KL} divergence between the discrete and the approximated posterior. In particular, we propose the factors in \EQ{aproxpriorMMSE} to be estimated iteratively. At iteration $\ell$ these factors are given by the Gaussian probabilities,
\begin{equation}
t_\iter^{[\ell]}(\beforechannel_\iter)\sim \cgauss{\beforechannel_\iter}{\mu_{t_\iter}^{[\ell]}}{\sigma_{t_\iter}^{2[\ell]}}.
\end{equation}
The minimization of the \ac{KL} divergence amounts to matching the moments,  
\begin{equation}\LABEQ{MM}
q_E^{[\ell]}(\beforechannel_\iter)p(\beforechannel_\iter) \stackrel{\mbox{\begin{tabular}{c}moment\\matching\end{tabular}}}{\longleftrightarrow} q_E^{[\ell]}(\beforechannel_\iter)t_\iter^{[\ell+1]}(\beforechannel_\iter)
\end{equation}
where $q_E^{[\ell]}(\beforechannel_\iter)$ is given by \EQ{extrinsic} with moments $\mu_{E_\iter}^{[\ell]}$ and $\sigma_{E_\iter}^{2[\ell]}$ and with $t_\iter(\beforechannel_\iter)$ replaced by $t_\iter^{[\ell]}(\beforechannel_\iter)$.
One iteration of this procedure is detailed in \ALG{MMD}. Since this algorithm suffers from instabilities and negative variances, we introduced
 a damping (with factor $\beta$), minimum allowed variance ($\epsilon$) and control of negative variances at every iteration. 

The full algorithm, described in \ALG{SmoothEP},  initializes in Step 1) the approximating factors, $t_\iter^{[1]}(\beforechannel_\iter)$, to the values given by the LMMSE, i.e., following \EQ{aproxpriorMMSE}. Then it runs \ALG{SmoothLMMSE} and \ALG{MMD} along $\iterep$ iterations to refine these terms. In \ALG{SmoothLMMSE}, using  $t_\iter^{[\ell]}(\beforechannel_\iter)$, we compute the full approximations $q^{[\ell]}(\vect{\state}_\iter)$, their marginals, $q^{[\ell]}({\beforechannel_\iter})$, and the extrinsic probabilities, $q_E^{[\ell]}({\beforechannel_\iter})$. Then we apply the moment matching in \ALG{MMD} to re-estimate the approximating factors as $t_\iter^{[\ell+1]}(\beforechannel_\iter)$. 

We denote this proposal as \ac{KSEP} turbo equalizer. \notaIb{Its computational complexity is $\order((\iterep+1)2\tamframe\ntaps^2)$, i.e., \notaI{the complexity of computing the Kalman smoother plus running $\iterep$ times the EP algorithm.}}
%
%
%
%
%
%
Note that the EP approach is applied once the smoothing step is performed. 

In the turbo equalization, we run  \ALG{SmoothEP} along $\nturbo+1$ iterations. The first time \ALG{SmoothEP}  is run, the inputs $p(\beforechannel_\iter)$ are initialized to uniform discrete values. Then, after every turbo iteration, they are set to the extrinsic probabilities provided by the channel decoder, $p_D(\beforechannel_\iter)$. Following the guidelines in \cite{Santos18}, we propose to set the parameters to: $\iterep=3$, $\beta=\min(\exp{(t/1.5)}/10,0.7)$, where $t\in[0,\nturbo]$ is the number of the current turbo iteration and $\epsilon=10^{-8}$.

\begin{algorithm}[!tb]
\begin{algorithmic}
\STATE 
{\bf Given inputs}: $p(\beforechannel_\iter)$, $t_\iter^{[\ell]}(\beforechannel_\iter)$ with moments $\mu_{t_\iter}^{[\ell]},\sigma_{t_\iter}^{2[\ell]}$ and $q_E^{[\ell]}(\beforechannel_\iter)$ with moments $\mu_{E_\iter}^{[\ell]},\sigma_{E_\iter}^{2[\ell]}$ 
\STATE
\vspace{0.2cm}
1) Compute the moments $\mu_{{\widehat{p}}_\iter}^{[\ell]},\sigma_{{\widehat{p}}_{\iter,aux}}^{2[\ell]}$ of the discrete posterior $\widehat{p}^{[\ell]}(\beforechannel_\iter)\;\propto\;q_E^{[\ell]}(\beforechannel_\iter)p(\beforechannel_\iter)$. Set a {\it minimum allowed variance}, \mbox{$\sigma_{{\widehat{p}}_\iter}^{2[\ell]}=\max(\epsilon,\sigma_{{\widehat{p}}_{\iter,aux}}^{2[\ell]})$}. 

\STATE
2)  Run \textit{moment matching}:  Set the mean and variance of the unnormalized Gaussian distribution
\begin{equation}\LABEQ{updategaussian}
q_E^{[\ell]}(\beforechannel_\iter)\cdot\cgauss{\beforechannel_\iter}{\mu_{t_\iter,new}^{[\ell{+1}]}}{\sigma_{t_\iter,new}^{2[\ell{+1}]}}
\end{equation}
equal to $\mu_{{\widehat{p}}_\iter}^{[\ell]}$ and $\sigma_{{\widehat{p}}_\iter}^{2[\ell]}$, to get the solution
\begin{align}
\sigma_{t_\iter,new}^{2[\ell+1]}&=\frac{\sigma_{{\widehat{p}}_\iter}^{2[\ell]}\sigma_{E_\iter}^{2[\ell]}}{\sigma_{E_\iter}^{2[\ell]}-\sigma_{{\widehat{p}}_\iter}^{2[\ell]}} , \LABEQ{Lambdak1new} \\
\mu_{t_\iter,new}^{[\ell+1]}&= \sigma_{t_\iter,new}^{2[\ell+1]}{\left( \frac{\mu_{{\widehat{p}}_\iter}^{[\ell]}}{\sigma_{{\widehat{p}}_\iter}^{2[\ell]}}-\frac{{\mu^{[\ell]}_{E_\iter}}}{\sigma_{E_\iter}^{2[\ell]}} \right)} . 
\end{align}
\STATE
3) Run \textit{damping}: Update the values as
\begin{align}
\sigma_{t_\iter}^{2[\ell+1]}&=\left(\beta\frac{1}{\sigma_{t_\iter,new}^{2[\ell+1]}} + (1-\beta)\frac{1}{\sigma_{t_\iter}^{2[\ell]}}\right)\inv \LABEQ{Lambdak1} ,\\
\mu_{t_\iter}^{[\ell+1]}&=\sigma_{t_\iter}^{2[\ell+1]}\left(\beta \frac{\mu_{t_\iter,new}^{[\ell+1]}}{\sigma_{t_\iter,new}^{2[\ell+1]}} + (1-\beta)\frac{\mu_{t_\iter}^{[\ell]}}{\sigma_{t_\iter}^{2[\ell]}}\right). \LABEQ{Lambdak2}
\end{align}
\STATE
4) Control of \textit{negative variances}:
\IF{${\sigma_{t_\iter}^{2[\ell+1]}}<0$}
\vspace{-0.3cm}
\STATE
\begin{align}
\sigma_{t_\iter}^{2[\ell+1]}=\sigma_{t_\iter}^{2[\ell]}, \,\,\,\,\,\,\,\, \mu_{t_\iter}^{[\ell+1]}=\mu_{t_\iter}^{[\ell]}. 
\end{align}
\ENDIF
\STATE
{\bf Output}: $\sigma_{t_\iter}^{2[\ell+1]}, \mu_{t_\iter}^{[\ell+1]}$
\end{algorithmic}
\caption{Moment Matching and Damping at Iteration $\ell$}\LABALG{MMD}
\end{algorithm}


\begin{algorithm}[!tb]
\begin{algorithmic}
\STATE 
{\bf Inputs}: $p(\beforechannel_\iter)$ for $\iter=1,...,\tamframe$
%
and $\afterchannel_\iter$ for $\iter=1,...,\tamframe+\memch$
\STATE
1) Initialization: compute $t_\iter^{[1]}(\beforechannel_\iter)=\mbox{Proj}_G[p(\beforechannel_\iter)]$. 
\FOR {$\ell=1,...,\iterep$}
\STATE
2) Run \ALG{SmoothLMMSE} with $t_\iter^{[\ell]}(\beforechannel_\iter)$ to obtain $q_E^{[\ell]}(\beforechannel_\iter)$, for $\iter=1,...,\tamframe$. \\
%
\FOR {$\iter=1,...,\tamframe$}
\STATE
3) Run \ALG{MMD} with $p(\beforechannel_\iter)$, $t_\iter^{[\ell]}(\beforechannel_\iter)$ and $q_E^{[\ell]}(\beforechannel_\iter)$ to obtain $\sigma_{t_\iter}^{2[\ell+1]}, \mu_{{t}_\iter}^{[\ell+1]}$. \\
\ENDFOR 
\ENDFOR 
\STATE
4) With the values $\mu_{t_\iter}^{[\iterep+1]}$ and $\sigma_{t_\iter}^{2[\iterep+1]}$ computed after EP, obtain $q_E^{[\iterep+1]}(\beforechannel_\iter)$ by running \ALG{SmoothLMMSE}.  
\STATE
{\bf Output}: Deliver $q_E^{[\iterep+1]}(\beforechannel_\iter)$ to the decoder  for $\iter=1,...,\tamframe$
\end{algorithmic}
\caption{Kalman Smoothing EP (KSEP) Equalizer}\LABALG{SmoothEP}
\end{algorithm}


%
%
%

\subsection{Discussion about EP-based proposals}


The \ac{BEP} equalizer is proposed in \cite{Santos18}, where the EP parameters are revised and optimized for turbo equalization. However, it requires to invert a matrix of size $\tamframe\times\tamframe$, yielding a quadratic complexity in the frame length. This complexity can be reduced by means of a smoothing implementation, such as \ac{BP-EP} \cite{Sun15} or \ac{SEP} \cite{Santos17}. The \ac{BP-EP} uses a window of size $\ntaps$ and applies the EP to better approximate with Gaussians the discrete information at the output of the channel decoder. On the other hand, the \ac{SEP} exhibits a much better performance than the \ac{BP-EP} in terms of BER but it has a higher computational complexity and it does not properly handle the information coming from the channel decoder. The \ac{SEP} assumes that the true discrete priors used during the moment matching procedure of the EP algorithm are uniform, even after the feedback from the channel decoder. It applies the EP over the forward and backward Kalman filters, i.e., the EP update in the \ac{SEP} is computed with just part of the observations and the performance is degraded. To overcome these problems, a window larger than the channel length, of size $2\ntaps-1$, is used. \notaIb{This yields a complexity of $\order((\iterep+1)\tamframe(2\ntaps-1)^3)$ for the forward/backward steps plus a complexity of $\order(\tamframe(2\ntaps-1)^3)$ due to the smoothing step. Since $S=10$ iterations of the moment matching algorithm are needed, the final computational complexity is of $\order(12\tamframe(2\ntaps-1)^3)$} \notaI{or, approximately, $\order(96\tamframe\ntaps^3)$ if we just keep the quadratic term in $\ntaps$. } 



In contrast, and as explained above, the novel KSEP proposal applies the EP at the smoothing level, where information from the whole observation vector is exploited. This fact allows to reduce the dimensions of the matrices to invert to be $\ntaps\times\ntaps$.  
Also, it is optimized for turbo equalization, setting the priors used in the moment matching to the non-uniform distributions provided by the channel decoder. The EP parameters are optimized for turbo equalization, needing only $S=3$ iterations of the moment matching to achieve convergence. This yields a complexity of $\order(\notaIb{8}\tamframe\ntaps^2)$, i.e., it is of the same order as the one of the BP-EP or the Kalman smoother, with $\order(\notaIb{2}\tamframe\ntaps^2)$. In\TAB{comp} we include a comparison in terms of complexity (per turbo loop) between the EP-based equalizers in \cite{Santos18,Santos17,Sun15} and our proposal.

\begin{table}[htb]
\begin{center}
\begin{tabular}{c c c c c}
\toprule
Algorithm & Complexity   \\
\midrule
Kalman Smoother &  $\order(\notaIb{2}\tamframe\ntaps^2)$   \\
\ac{BEP} \cite{Santos18} &  {$\order(4\tamframe^2\ntaps)$}   \\
\ac{BP-EP} {\cite{Sun15}} & $\order(\notaIb{2}\tamframe\ntaps^2)$   \\
\ac{SEP} {\cite{Santos17}} & 
$\notaI{\order(96\tamframe\ntaps^3)}$   \\
\ac{KSEP} & $\order(\notaIb{8}\tamframe\ntaps^2)$   \\
\bottomrule
\end{tabular}
\captionof{table}{\small Complexity comparison between EP-based equalizers per turbo iteration. }\LABTAB{comp}
\end{center}
\end{table}

\section{Experimental results}\LABSEC{results}

In this section we compare the performance in terms of \ac{BER} of our proposal, \ac{KSEP}, with the block LMMSE algorithm 
and other EP-based proposals found in the literature, such as the \ac{BEP} \cite{Santos18}, \ac{SEP} \cite{Santos17} and \ac{BP-EP} \cite{Sun15}. We use different modulations and lengths for the channel. The results are averaged over $100$ random channels and $10^4$ random encoded words of length $\t=4096$ (per channel realization). The parameters of the KSEP are set to the same values as given in \cite{Santos18} for the \ac{BEP}: $\iterep=3$, $\nturbo=5$ turbo iterations, $\beta=\min(\exp{(t/1.5)}/10,0.7)$, where $t\in[0,\nturbo]$ is the number of the current turbo iteration and $\epsilon=10^{-8}$. 
Each channel tap is independent and identically Gaussian distributed 
with zero mean and variance equal to $1/{\ntaps}$. The absolute value of LLRs given to the decoder is limited to $5$ in order to avoid very confident probabilities. 
A (3,6)-regular \ac{LDPC} of rate $1/2$ is used, decoded with a belief propagation algorithm with a maximum of $100$ iterations.  

In \FIG{4PAM5taps} we show the \ac{BER} for a $4$-PAM constellation and random channels of $\ntaps=5$ real-valued taps. As can be observed, the BP-EP \cite{Sun15} exhibits the highest error in terms of \ac{BER}, even compared to the LMMSE. Note that the BP-EP does not properly control the negative variances and does not include any damping procedure or minimum allowed variance. 
The \ac{BEP} and \ac{KSEP} share the same performance. These two proposals have gains close to 2 dB with respect to the LMMSE and 2 dB with respect to \ac{BP-EP}. The performance of the \ac{SEP} degrades when comparing with \ac{BEP} and \ac{KSEP}. Although the SEP is not optimized for turbo equalization it improves the LMMSE in 0.75 dB. 

\begin{figure}[t!]
\scalebox{0.9}{
%
%
%
%
\begin{tikzpicture}[scale=1]

\begin{axis}[%
width=2.9in,
height=2in,
scale only axis,
every axis/.append style={font=\small},
xmajorgrids,
xmin=6,
xmax=15,
xlabel style={align=center}, xlabel={$\EbNo$ (dB)},
ymode=log,
ymin=1e-05,
ymax=1e-1,
yminorticks=true,
ylabel={BER},
ymajorgrids,
yminorgrids,
legend style={at={(1,1)},anchor=north east,draw=black,fill=white,legend cell align=left,
,font=\footnotesize
}
]

\addplot [color=blue,solid,mark=triangle,mark size=2.2,line width=1.1,mark options={solid,,rotate=180}]
  table[row sep=crcr]{
5	0.0949586595052632	\\
6	0.0536083041157895	\\
7	0.0315994346736842	\\
8	0.0177573640210526	\\
9	0.00940267406315789	\\
10	0.00498847324210526	\\
11	0.00252265511578947	\\
12	0.000789576821052631	\\
13	0.000136677557894737	\\
14	1.38364526315789e-05	\\
15	1.41863157894737e-06	\\
};
\addlegendentry{LMMSE};

\addplot [color=mycolor3,solid,mark=o,mark size=2.2,line width=1.1,mark options={solid}]
  table[row sep=crcr]{
6	0.0276323038421053	\\
7	0.0133158636842105	\\
8	0.00675374443157895	\\
9	0.00313419002105263	\\
10	0.00117506323157895	\\
11	0.000270065842105263	\\
12	3.33677052631579e-05	\\
13	0	\\
14	0	\\
};
\addlegendentry{BEP \cite{Santos18}};

\addplot [color=mycolor1,solid,mark=+,mark size=2.2,line width=1.1,mark options={solid}]
  table[row sep=crcr]{
5	0.0717526718421053	\\
6	0.0400711205894737	\\
7	0.0254280713368421	\\
8	0.015767546	\\
9	0.00917441421052631	\\
10	0.00501468063157895	\\
11	0.0030275452631579	\\
12	0.00129220747368421	\\
13	0.000341920736842105	\\
14	4.03987368421053e-05	\\
15	4.05021052631579e-06	\\
};
\addlegendentry{BP-EP \cite{Sun15}};

\addplot [color=mycolor,solid,mark=diamond,mark size=2.2,line width=1.1,mark options={solid}]
  table[row sep=crcr]{
6	0.0493853827473684	\\
7	0.0274411990526316	\\
8	0.0143541842210526	\\
9	0.00715678668526316	\\
10	0.00377699283157895	\\
11	0.00162455167368421	\\
12	0.000288281147368421	\\
13	1.56864736842105e-05	\\
14	0	\\
};
\addlegendentry{SEP \cite{Santos17}};

\addplot [color=mycolor2,solid,mark=x,mark size=2.6,line width=1.1,mark options={solid}]
  table[row sep=crcr]{
6	0.0276348371157895	\\
7	0.0133062011936842	\\
8	0.00674835982105263	\\
9	0.00315526582105263	\\
10	0.00117261464210526	\\
11	0.000253084115789474	\\
12	3.17640947368421e-05	\\
13	0	\\
14	0	\\
};
\addlegendentry{KSEP};

\end{axis}

\end{tikzpicture}%
}
\caption{\small BER along $\EbNo$ for \ac{BEP} (\textcolor{mycolor3}{$\circ$}) \cite{Santos18}, \ac{SEP} (\textcolor{mycolor}{$\diamond$})  \cite{Santos17}, \ac{KSEP}  (\textcolor{mycolor2}{$\times$}), \ac{BP-EP} (\textcolor{mycolor1}{$+$}) \cite{Sun15} and LMMSE (\textcolor{blue}{$\triangledown$}) turbo equalizers, $4$-PAM and averaged over 100 random channels with $\ntaps=5$ real taps after $\nturbo=5$ turbo iterations. }
\LABFIG{4PAM5taps}
\end{figure}

In \FIG{64QAM7taps} we increase the order of the constellation and the length of the channel. In particular, we show the \ac{BER} for random channels of $\ntaps=7$ complex-valued taps and a $64$-QAM constellation. We decided not to include the \ac{BP-EP} since, as discussed in the previous experiment in \FIG{4PAM5taps}, its results degrade when using multilevel modulations. We plotted the performance after different number of turbo iterations. Specifically, for standalone equalization (a) and for turbo equalization after $\nturbo=2$ (b) and $\nturbo=5$ (c) iterations. We observe that in these three cases the \ac{KSEP} and \ac{BEP} have the same performance. In \FIG{64QAM7taps} (a), it can be seen that SEP is slightly better than BEP and KSEP. The reason is that it is run with $\iterep=10$ iterations, while BEP and KSEP reduce it to $\iterep=3$ since they have been designed for turbo equalization. 
The KSEP achieves the performance of the SEP in  \FIG{64QAM7taps}  (a) by just increasing its number of EP iterations to $\iterep=6$.  In \FIG{64QAM7taps} (b), where the BER after $\nturbo=2$ turbo iterations is depicted, both BEP and KSEP quite outperform the SEP performance. \notaIb{The reason is that, in addition to exploiting the whole vector of observations when applying EP -in contrast to SEP-, they are properly handling the discrete information returned by the channel decoder.}
Specifically, they have gains of 3 dB and 4.5 dB  with respect to the SEP and LMMSE, respectively. In \FIG{64QAM7taps} (c), the \ac{SEP} shows better performance than the LMMSE but it is quite far from \ac{KSEP} and \ac{BEP}. Specifically, \ac{KSEP} and \ac{BEP} improve  the performance of \ac{SEP} in 4 dB. 

In \FIG{128QAM7taps} we reproduce the scenario in \FIG{64QAM7taps} with a higher order modulation, a $128$-QAM. Again, for the standalone equalization case showed in \FIG{128QAM7taps} (a), the performance of \ac{SEP} is slightly better than the one of \ac{KSEP}/\ac{BEP}, although it has been checked that they achieve the SEP performance by increasing the number of iterations to $\iterep=6$. 
For the turbo equalization case in \FIG{128QAM7taps} (b)-(c), the \ac{SEP} improves the LMMSE but its performance is quite far from the one of the \ac{KSEP} and \ac{BEP}. Specifically, \ac{KSEP} and \ac{BEP} have gains of 5 dB  in comparison with the LMMSE in \FIG{128QAM7taps} (b) and gains of almost 6 dB in \FIG{128QAM7taps} (c). Similarly to \FIG{4PAM5taps} and \FIG{64QAM7taps}, the \ac{SEP} is in between \ac{BEP}/\ac{KSEP} and \ac{LMMSE}.

%

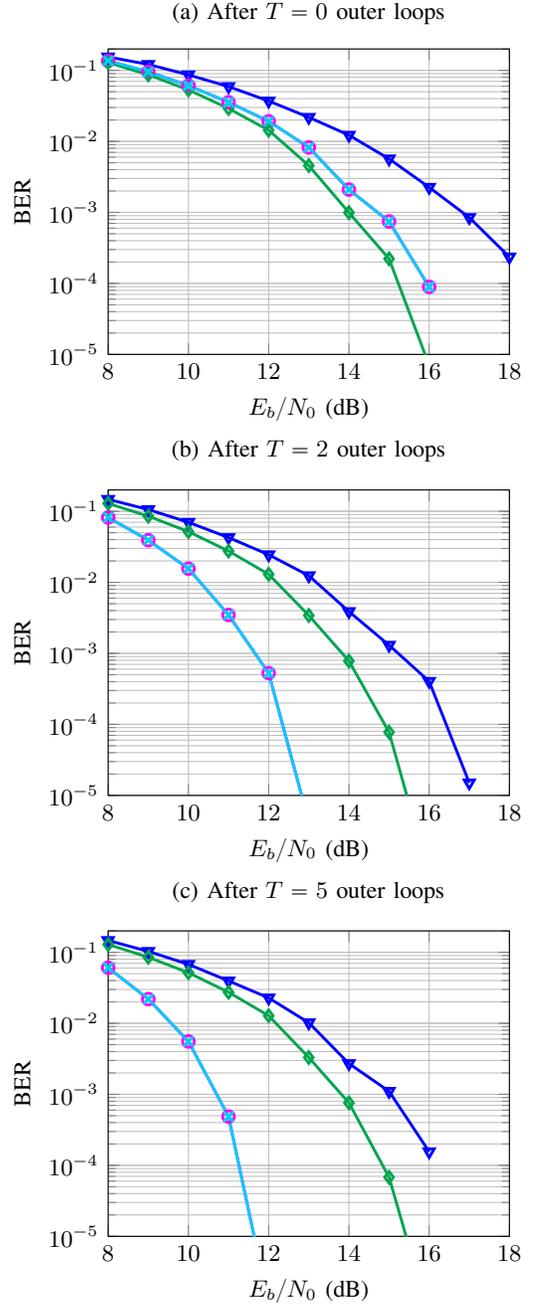
\begin{figure}[htb]
\centering
\scalebox{1}{
%
%
%
%
%
\begin{tikzpicture}[scale=1]




\begin{axis}[%
width=2.1in,
height=1.6in,
scale only axis,
every axis/.append style={font=\small},
xmajorgrids,
xmin=8,
xmax=18,
xlabel style={align=center}, xlabel={$\EbNo$ (dB)},
ymode=log,
ymin=1e-05,
ymax=2e-1,
yminorticks=true,
ylabel={BER},
name=plot1,
ymajorgrids,
yminorgrids,
title={(a) After $\nturbo=0$ outer loops},
legend style={at={(1,1)},anchor=north east,draw=black,fill=white,legend cell align=left
,font=\footnotesize
}
]

\addplot [color=blue,solid,mark=triangle,mark size=2.2,line width=1.1,mark options={solid,,rotate=180}]
  table[row sep=crcr]{
8	0.154459904642105	\\
9	0.121247725111579	\\
10	0.0864169155652632	\\
11	0.0591107596105263	\\
12	0.0370325256768421	\\
13	0.0218259012345263	\\
14	0.0122313132164211	\\
15	0.00567419342947369	\\
16	0.00225513909473684	\\
17	0.000847352842105263	\\
18	0.0002357888	\\
};

\addplot [color=mycolor3,solid,mark=o,mark size=2.2,line width=1.1,mark options={solid}]
  table[row sep=crcr]{
8	0.136029115226316	\\
9	0.0954820079136842	\\
10	0.0611710907136842	\\
11	0.0355691905589474	\\
12	0.0192050123894737	\\
13	0.00821268852926316	\\
14	0.00209635222947368	\\
15	0.000743986642105263	\\
16	8.92776842105263e-05	\\
17	0	\\
18	0	\\
};

\addplot [color=mycolor,solid,mark=diamond,mark size=2.2,line width=1.1,mark options={solid}]
  table[row sep=crcr]{
8	0.128749659014526	\\
9	0.08650659064	\\
10	0.0531566095063158	\\
11	0.0290445077052632	\\
12	0.0143530968989474	\\
13	0.00456243444210526	\\
14	0.00099347795031579	\\
15	0.000222204157894737	\\
16	7.05183157894737e-06	\\
};

\addplot [color=mycolor2,solid,mark=x,mark size=2.6,line width=1.1,mark options={solid}]
  table[row sep=crcr]{
8	0.136029092573684	\\
9	0.0954819920926316	\\
10	0.0611709280736842	\\
11	0.0355690591863158	\\
12	0.0192050584210526	\\
13	0.0082127986871579	\\
14	0.00209628582947369	\\
15	0.000743991894736842	\\
16	8.92776842105263e-05	\\
17	0	\\
18	0	\\
};

\end{axis}

\begin{axis}[%
width=2.1in,
height=1.6in,
scale only axis,
every axis/.append style={font=\small},
xmajorgrids,
xmin=8,
xmax=18,
xlabel style={align=center}, xlabel={$\EbNo$ (dB) },
ymode=log,
ymin=1e-05,
ymax=2e-1,
yminorticks=true,
ylabel={BER},
name=plot2,
at=(plot1.below south west), anchor=above north west,
ymajorgrids,
yminorgrids,
title={\begin{tabular}{c} (b) After $\nturbo=2$ outer loops \end{tabular}},
legend style={at={(1,1)},anchor=north east,draw=black,fill=white,legend cell align=left,
,font=\footnotesize
}
]

\addplot [color=blue,solid,mark=triangle,mark size=2.2,line width=1.1,mark options={solid,,rotate=180}]
  table[row sep=crcr]{
8	0.147489931589474	\\
9	0.105850946821053	\\
10	0.0700773083694737	\\
11	0.0427165658421053	\\
12	0.0244927029578947	\\
13	0.0124164852210526	\\
14	0.00387948891578948	\\
15	0.00130380149473684	\\
16	0.000403618852631579	\\
17	1.50236210526316e-05	\\
18	0	\\
};

\addplot [color=mycolor3,solid,mark=o,mark size=2.2,line width=1.1,mark options={solid}]
  table[row sep=crcr]{
8	0.0818417521084211	\\
9	0.0392314143284211	\\
10	0.0155824633473684	\\
11	0.00347370464505263	\\
12	0.000526577968421053	\\
13	4.26090526315789e-06	\\
14	7.60694736842105e-07	\\
15	0	\\
16	0	\\
17	0	\\
18	0	\\
};

\addplot [color=mycolor,solid,mark=diamond,mark size=2.2,line width=1.1,mark options={solid}]
  table[row sep=crcr]{
8	0.128986954642105	\\
9	0.0853604988	\\
10	0.0517232875094737	\\
11	0.0275798616421053	\\
12	0.0129807861452632	\\
13	0.0034180712968421	\\
14	0.000779836947368421	\\
15	7.76676115789473e-05	\\
16	7.55547368421053e-07	\\
};

\addplot [color=mycolor2,solid,mark=x,mark size=2.6,line width=1.1,mark options={solid}]
  table[row sep=crcr]{
8	0.081841809332421	\\
9	0.0392315355484211	\\
10	0.0155823127052632	\\
11	0.00347399283747368	\\
12	0.000526609115789474	\\
13	4.26090526315789e-06	\\
14	7.60694736842105e-07	\\
15	0	\\
16	0	\\
17	0	\\
18	0	\\
};

\end{axis}

\begin{axis}[%
width=2.1in,
height=1.6in,
scale only axis,
every axis/.append style={font=\small},
xmajorgrids,
xmin=8,
xmax=18,
xlabel style={align=center}, xlabel={$\EbNo$ (dB) },
ymode=log,
ymin=1e-05,
ymax=2e-1,
yminorticks=true,
ylabel={BER},
name=plot3,
at=(plot2.below south west), anchor=above north west,
ymajorgrids,
yminorgrids,
title={\begin{tabular}{c} (c) After $\nturbo=5$ outer loops \end{tabular}},
legend style={at={(1,1)},anchor=north east,draw=black,fill=white,legend cell align=left,
,font=\footnotesize
}
]

\addplot [color=blue,solid,mark=triangle,mark size=2.2,line width=1.1,mark options={solid,,rotate=180}]
  table[row sep=crcr]{
8	0.146659008189474	\\
9	0.102988026610526	\\
10	0.0676474062736842	\\
11	0.0395697452315789	\\
12	0.022788072368421	\\
13	0.0101980994105263	\\
14	0.00271274694736842	\\
15	0.00110355967368421	\\
16	0.000155432536842105	\\
17	0	\\
18	0	\\
};

\addplot [color=mycolor3,solid,mark=o,mark size=2.2,line width=1.1,mark options={solid}]
  table[row sep=crcr]{
8	0.0606533684421053	\\
9	0.0219154082105263	\\
10	0.00555121468421053	\\
11	0.000485989305263158	\\
12	1.11021052631579e-06	\\
13	0	\\
14	0	\\
15	0	\\
16	0	\\
17	0	\\
18	0	\\
};

\addplot [color=mycolor,solid,mark=diamond,mark size=2.2,line width=1.1,mark options={solid}]
  table[row sep=crcr]{
8	0.128895138115789	\\
9	0.0851459117368421	\\
10	0.0514852824105263	\\
11	0.0273811412421053	\\
12	0.0127900634631579	\\
13	0.00329309613684211	\\
14	0.000759036042105263	\\
15	6.7963747368421e-05	\\
16	7.50410526315789e-07	\\
};

\addplot [color=mycolor2,solid,mark=x,mark size=2.6,line width=1.1,mark options={solid}]
  table[row sep=crcr]{
8	0.0606545852557895	\\
9	0.0219133983684211	\\
10	0.00555232694736842	\\
11	0.000485701568421053	\\
12	1.11021052631579e-06	\\
13	0	\\
14	0	\\
15	0	\\
16	0	\\
17	0	\\
18	0	\\
};

\end{axis}

\end{tikzpicture}%
}
\caption{\small BER along $\EbNo$ for for \ac{BEP} (\textcolor{mycolor3}{$\circ$}) \cite{Santos18}, \ac{SEP} (\textcolor{mycolor}{$\diamond$}) \cite{Santos17}, \ac{KSEP} (\textcolor{mycolor2}{$\times$}) and LMMSE  (\textcolor{blue}{$\triangledown$}) turbo equalizers, $64$-QAM and averaged over 100 random channels with $\ntaps=7$ complex taps.}\LABFIG{64QAM7taps}
\end{figure} 

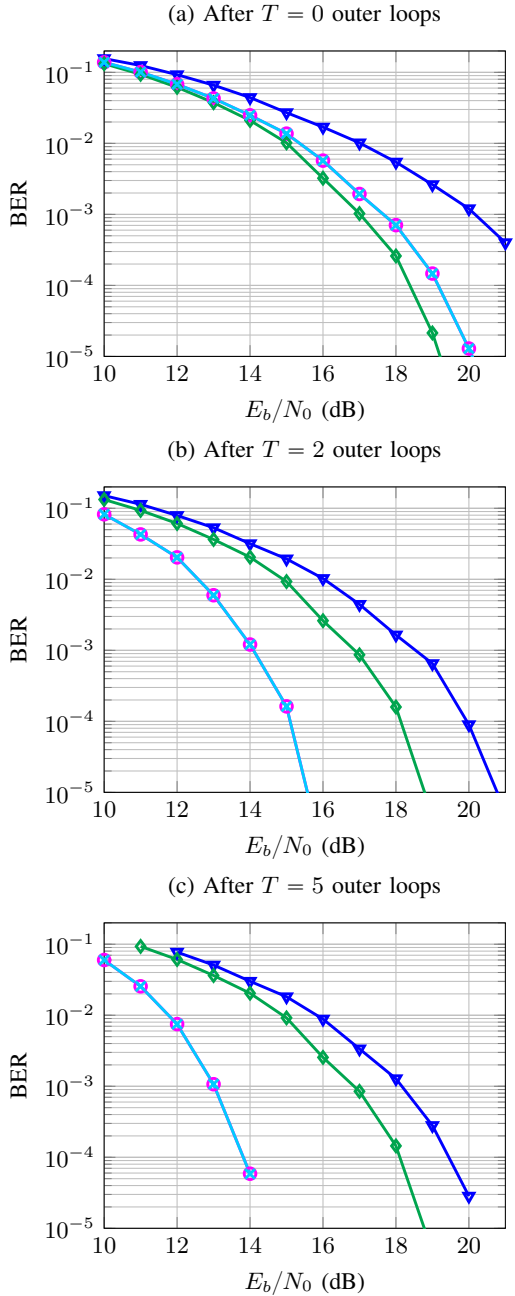
\begin{figure}[htb]
\centering
\scalebox{1}{
%
%
%
%
%
\begin{tikzpicture}[scale=1]




\begin{axis}[%
width=2.1in,
height=1.6in,
scale only axis,
every axis/.append style={font=\small},
xmajorgrids,
xmin=10,
xmax=21,
xlabel style={align=center}, xlabel={$\EbNo$ (dB)},
ymode=log,
ymin=1e-05,
ymax=2e-1,
yminorticks=true,
ylabel={BER},
name=plot1,
ymajorgrids,
yminorgrids,
title={(a) After $\nturbo=0$ outer loops},
legend style={at={(1,1)},anchor=north east,draw=black,fill=white,legend cell align=left
,font=\footnotesize
}
]

\addplot [color=blue,solid,mark=triangle,mark size=2.2,line width=1.1,mark options={solid,,rotate=180}]
  table[row sep=crcr]{
10	0.156267072010526	\\
11	0.124811063684211	\\
12	0.0926877078421052	\\
13	0.0660103145292631	\\
14	0.0442095232736842	\\
15	0.0271008486526316	\\
16	0.0169534647684211	\\
17	0.0101693326968421	\\
18	0.00545503371578947	\\
19	0.0026247296	\\
20	0.00120238893684211	\\
21	0.000399814870526316	\\
};

\addplot [color=mycolor3,solid,mark=o,mark size=2.2,line width=1.1,mark options={solid}]
  table[row sep=crcr]{
10	0.138370602221053	\\
11	0.100816354442105	\\
12	0.0678921970915789	\\
13	0.0429870631924211	\\
14	0.0248431010930526	\\
15	0.0137143796168421	\\
16	0.00571388945368421	\\
17	0.00193796618947368	\\
18	0.00070578738	\\
19	0.000146869357894737	\\
20	1.29164315789474e-05	\\
21	7.70968421052632e-07	\\
};

\addplot [color=mycolor,solid,mark=diamond,mark size=2.2,line width=1.1,mark options={solid}]
  table[row sep=crcr]{
10	0.132240897550526	\\
11	0.0937927874	\\
12	0.0615727385957895	\\
13	0.0374344828147368	\\
14	0.0210383918084211	\\
15	0.0101798468631579	\\
16	0.0032300762	\\
17	0.00102648118210526	\\
18	0.000259698684210526	\\
19	2.13969715789474e-05	\\
20	5.75652631578947e-07	\\
};

\addplot [color=mycolor2,solid,mark=x,mark size=2.6,line width=1.1,mark options={solid}]
  table[row sep=crcr]{
10	0.138370608568421	\\
11	0.1008163784	\\
12	0.0678925039336842	\\
13	0.0429869749345263	\\
14	0.0248434950526316	\\
15	0.0137143903263158	\\
16	0.00571400803263158	\\
17	0.00193785194736842	\\
18	0.000705694863157895	\\
19	0.000146895147368421	\\
20	1.29214842105263e-05	\\
21	7.70968421052632e-07	\\
};

\end{axis}

\begin{axis}[%
width=2.1in,
height=1.6in,
scale only axis,
every axis/.append style={font=\small},
xmajorgrids,
xmin=10,
xmax=21,
xlabel style={align=center}, xlabel={$\EbNo$ (dB) },
ymode=log,
ymin=1e-05,
ymax=2e-1,
yminorticks=true,
ylabel={BER},
name=plot2,
at=(plot1.below south west), anchor=above north west,
ymajorgrids,
yminorgrids,
title={(b) After $\nturbo=2$ outer loops },
legend style={at={(1,1)},anchor=north east,draw=black,fill=white,legend cell align=left,
,font=\footnotesize
}
]

\addplot [color=blue,solid,mark=triangle,mark size=2.2,line width=1.1,mark options={solid,,rotate=180}]
  table[row sep=crcr]{
10	0.150951079589474	\\
11	0.113734333715789	\\
12	0.0792170109073684	\\
13	0.0532558435578948	\\
14	0.0317554462505263	\\
15	0.0193938651789474	\\
16	0.0102712165473684	\\
17	0.00444267206315789	\\
18	0.00164715045263158	\\
19	0.000651279421052632	\\
20	8.97718947368421e-05	\\
21	5.32477894736842e-06	\\
};

\addplot [color=mycolor3,solid,mark=o,mark size=2.2,line width=1.1,mark options={solid}]
  table[row sep=crcr]{
10	0.0823586288294737	\\
11	0.0428810771713684	\\
12	0.0203621384842105	\\
13	0.0059667968	\\
14	0.00120826644210526	\\
15	0.000162505115789474	\\
16	1.26442105263158e-06	\\
17	0	\\
18	0	\\
19	0	\\
20	0	\\
21	0	\\
};

\addplot [color=mycolor,solid,mark=diamond,mark size=2.2,line width=1.1,mark options={solid}]
  table[row sep=crcr]{
10	0.132884592273684	\\
11	0.0933575388831579	\\
12	0.0611282442736842	\\
13	0.0364171772210526	\\
14	0.0205013953052632	\\
15	0.00929551735789474	\\
16	0.00261309051578947	\\
17	0.000864473526315789	\\
18	0.000159154031578947	\\
19	4.72861052631579e-06	\\
20	0	\\
};

\addplot [color=mycolor2,solid,mark=x,mark size=2.6,line width=1.1,mark options={solid}]
  table[row sep=crcr]{
10	0.0823582511242105	\\
11	0.0428812911532632	\\
12	0.0203618814210526	\\
13	0.00596680207368421	\\
14	0.00120829221052632	\\
15	0.000162433326315789	\\
16	1.26442105263158e-06	\\
17	0	\\
18	0	\\
19	0	\\
20	0	\\
21	0	\\
};

\end{axis}

\begin{axis}[%
width=2.1in,
height=1.6in,
scale only axis,
every axis/.append style={font=\small},
xmajorgrids,
xmin=10,
xmax=21,
xlabel style={align=center}, xlabel={$\EbNo$ (dB) },
ymode=log,
ymin=1e-05,
ymax=2e-1,
yminorticks=true,
ylabel={BER},
name=plot3,
at=(plot2.below south west), anchor=above north west,
ymajorgrids,
yminorgrids,
title={(c) After $\nturbo=5$ outer loops },
legend style={at={(1,1)},anchor=north east,draw=black,fill=white,legend cell align=left,
,font=\footnotesize
}
]

\addplot [color=blue,solid,mark=triangle,mark size=2.2,line width=1.1,mark options={solid,,rotate=180}]
  table[row sep=crcr]{
12	0.0773472572631579	\\
13	0.050897339631579	\\
14	0.0300951563157895	\\
15	0.0182232487368421	\\
16	0.00883577326315789	\\
17	0.00336328631578947	\\
18	0.00127845742105263	\\
19	0.000280248263157895	\\
20	2.83973157894737e-05	\\
21	0	\\
};

\addplot [color=mycolor3,solid,mark=o,mark size=2.2,line width=1.1,mark options={solid}]
  table[row sep=crcr]{
10	0.0599864581578947	\\
11	0.0255518218421053	\\
12	0.007501984	\\
13	0.00106921326315789	\\
14	5.87225789473684e-05	\\
15	0	\\
16	0	\\
17	0	\\
18	0	\\
19	0	\\
20	0	\\
21	0	\\
};

\addplot [color=mycolor,solid,mark=diamond,mark size=2.2,line width=1.1,mark options={solid}]
  table[row sep=crcr]{
11	0.0932203093473684	\\
12	0.0610064053052632	\\
13	0.0362752147578947	\\
14	0.0204182285473684	\\
15	0.00918697478947369	\\
16	0.00255032711578947	\\
17	0.000848828694736842	\\
18	0.000144659642105263	\\
19	4.6618e-06	\\
20	0	\\
};

\addplot [color=mycolor2,solid,mark=x,mark size=2.6,line width=1.1,mark options={solid}]
  table[row sep=crcr]{
10	0.059993421	\\
11	0.0255530423157895	\\
12	0.00750010263157895	\\
13	0.00107224263157895	\\
14	5.83885789473684e-05	\\
15	0	\\
16	0	\\
17	0	\\
18	0	\\
19	0	\\
20	0	\\
21	0	\\
};

\end{axis}

\end{tikzpicture}%
}
\caption{\small BER along $\EbNo$ for for \ac{BEP} (\textcolor{mycolor3}{$\circ$}) \cite{Santos18}, \ac{SEP} (\textcolor{mycolor}{$\diamond$}) \cite{Santos17}, \ac{KSEP} (\textcolor{mycolor2}{$\times$}) and LMMSE  (\textcolor{blue}{$\triangledown$}) turbo equalizers, $128$-QAM and averaged over 100 random channels with $\ntaps=7$ complex taps.}\LABFIG{128QAM7taps}
\end{figure} 


\section{Conclusions}\LABSEC{conc}

In this paper, we propose a novel Kalman smoothing EP-based equalizer, the \ac{KSEP}, that outperforms previous smoothing proposals found in the literature \cite{Santos17,Sun15}. The \ac{KSEP} first runs a forward and backward Kalman filter and then merges both approximations into a smoothing one, where the EP algorithm is applied. 
%
It solves the problems of previously proposed EP-based equalizers. Firstly, it avoids the degradation in the performance for high-order modulations of the BP-EP \cite{Sun15} approach. \notaIb{Secondly, in contrast to SEP, it applies EP at the smoothing level exploiting information from the whole observation vector and yielding better convergence properties. }Thirdly, rather than setting uniform priors in the moment matching procedure as the SEP, it properly handles the information coming from the channel decoder within the moment matching procedure. Also, it reduces the size of the matrices involved in the estimations in comparison to SEP therefore reducing the computational complexity. As illustrated in the experimental section, we have a remarkable gain in performance compared to previous equalizers based in EP and smoothing. The KSEP achieves the same performance as its block counterpart, the \ac{BEP}, as the Kalman smoother exhibits the same performance as the block LMMSE. Besides, the computational complexity yields $\order(\notaIb{8}\tamframe\ntaps^2)$, i.e., it is of the same order as the one of the BP-EP or the Kalman smoother, with $\order(\notaIb{2}\tamframe\ntaps^2)$. 

\appendix[Proof of \EQ{trueSmooth1}]\LABAPEN{Ap}
%
%


From \EQ{pugiveny}, the exact posterior probability distribution is
\begin{align}\LABEQ{realposterior}
&p(\vect{\beforechannel}|\vect{\afterchannel}) \;\propto\; p(\vect{\afterchannel}|\vect{\beforechannel})p(\vect{\beforechannel})
=\prod_{\iter=1 }^{\tamframe+\memch} p(\afterchannel_{\iter}|\vect{\state}_\iter)\prod_{\iter=-\memch+1}^{\tamframe+\memch}p(\beforechannel_{\iter}), 
\end{align}
 We aim at computing $p(\vect{\state}_k|\vect{\afterchannel})$ by using $p^F(\vect{\state}_k|\vect{\afterchannel}_{1:\iter})$ from the forward step and $p^B(\vect{\state}_k|\vect{\afterchannel}_{\iter:\tamframe+\memch})$ from the backward step. Accordingly, we split \EQ{realposterior} into two parts: received symbols until time $\iter$ and after time $\iter$,
\begin{align}\LABEQ{realposteriorsepbyk}
p(\vect{\beforechannel}|\vect{\afterchannel}) &\;\propto \; p(\vectcomp{\afterchannel}{1}{\iter}|\vectcomp{\beforechannel}{-\memch+1}{\iter}) p(\vectcomp{\beforechannel}{-\memch+1}{\iter})  \prod_{\iterw=\iter+1}^{\tamframe+\memch} p(\afterchannel_{\iterw}|\vect{\state}_\iterw)p(\beforechannel_{\iterw})\; \nonumber\\
&\;\propto\;{p(\vectcomp{\beforechannel}{-\memch+1}{\iter}|\vectcomp{\afterchannel}{1}{\iter})}{\prod_{\iterw=\iter+1}^{\tamframe+\memch} p(\afterchannel_{\iterw}|\vect{\state}_\iterw)p(\beforechannel_{\iterw})}.
\end{align}
We now marginalize $p(\vect{\beforechannel}|\vect{\afterchannel})$ over $\vectcomp{\beforechannel}{-\memch+1}{\iter-\memch-1}$ simplifying the first factor, from the forward description in \EQ{posState}, 
\begin{align}\LABEQ{realpostfor}
&p(\vectcomp{\beforechannel}{\iter-\memch}{\tamframe+\memch}|\vect{\afterchannel}) 
\;\propto\; p(\vect{\state}_k|\vect{\afterchannel}_{1:\iter}){\prod_{\iterw=\iter+1}^{\tamframe+\memch} p(\afterchannel_{\iterw}|\vect{\state}_\iterw)p(\beforechannel_{\iterw})}.
\end{align}
{To rewrite the last term of \EQ{realpostfor} into the posterior $p(\vectcomp{\beforechannel}{\iter-\memch}{\tamframe+\memch}|\vectcomp{\afterchannel}{\iter}{\tamframe+\memch})$, the following factors are missing
\begin{align}\LABEQ{factor}
p(\afterchannel_\iter|\vect{\state}_\iter)\prod_{\iterw=\iter-\memch}^{\iter}p(\beforechannel_\iterw).
\end{align}}
By multiplying and dividing \EQ{realpostfor} by \EQ{factor}, 
we get the equivalent distribution 
\begin{align}\LABEQ{aux1}
p(\vectcomp{\beforechannel}{\iter-\memch}{\tamframe+\memch}|\vect{\afterchannel}) \;\propto\; \frac{p(\vect{\state}_k|\vect{\afterchannel}_{1:\iter})p(\vectcomp{\beforechannel}{\iter-\memch}{\tamframe+\memch}|\vectcomp{\afterchannel}{\iter}{\tamframe+\memch})}{p(\afterchannel_\iter|\vect{\state}_\iter)\prod\limits_{\iterw=\iter-\memch}^{\iter}p(\beforechannel_\iterw)}.
\end{align}
Then after marginalizing \EQ{aux1} over the last symbols $\vectcomp{\beforechannel}{\iter+1}{\tamframe+\memch}$, it yields 
\begin{align}\LABEQ{trueSmooth}
p(\vect{\state}_k|\vect{\afterchannel}) \;\propto\; \frac{p(\vect{\state}_k|\vect{\afterchannel}_{1:\iter})p(\vect{\state}_k|\vect{\afterchannel}_{\iter:\tamframe+\memch})}{p(\afterchannel_\iter|\vect{\state}_\iter)\prod\limits_{\iterw=\iter-\memch}^{\iter}p(\beforechannel_\iterw)}.
\end{align}
%

%
%

\ifCLASSOPTIONcaptionsoff
  \newpage
\fi



%
%
%

\bibliographystyle{IEEEtran}
\bibliography{allBib}

%

%
%
%




\end{document}